# Measurements of the sticking time and sticking probability of Rb atoms on a polydimethylsiloxane coating

Revised 5/01/2014

S.N. Atutov*, A.I. Plekhanov

*Institute of Automation and Electrometry of the Siberian Branch of the Russian Academy of Science, Koptyug Ave. 1, Novosibirsk, Russia, 630090*
*Novosibirsk State University, Pirogova Str., 2, Novosibirsk, Russia, 630090*

We present the results of a systematic study of Knudsen's flow of Rb atoms in cylindrical capillary cells coated with a polydimethylsiloxane (PDMS) compound. The purpose of the investigation is to determine the characterization of the coating in terms of the sticking probability and sticking time of Rb on the two types of coating of high and median viscosities. We report the measurement of the sticking probability of the Rb atom to the coating that is equal to $4{,}3 \cdot 10^{-5}$ and corresponds to the number of bounces of $2{,}3 \cdot 10^4$ at room temperature. These parameters are the same for the two kinds of PDMS used. We find that at room temperature the sticking times for high viscosity and median viscosity PDMS are $22\pm3$ μs and $49\pm6$ μs, respectively. These sticking times are about million times larger than the sticking time derived from the surface Rb atom adsorption energy and temperature of the coating. A tentative explanation of this surprising result, based on the bulk diffusion of the atoms that collide with the surface and penetrate inside of the coating, is proposed. The results can be important to many resonance cell experiments, such as the efficient magneto-optical trapping of rare elements or radioactive isotopes and in experiments on the light-induced drift effect.

**PACS:** 34.35.+a, 68.43.Mn, 34.20.Mq, 34.90.+q,

## I. INTRODUCTION

The major technical difficulty in many resonance cell experiments, such as cooling and trapping short-lived radioactive isotopes using a magneto-optical trap (MOT) [1-10] or experiments on light-induced drift (LID) [11-19] lies in the atomic vapor interaction with the inner wall of the resonance cell.

Since short-lived radioactive isotopes are available only in limited quantities, an efficient optical trapping process is of great importance to the possibility to create large samples of these elements. An improvement in the collection efficiency of an MOT is a key consideration in experiments that use very weak atomic fluxes. Often in this kind of experiment, the MOT cell is connected to a source of radioactive ions through the entrance port to a beam transport line. Radioactive ions from a beam transport line are injected into the cell and impinge on the neutralizer (a hot metal plate placed at the far end of the cell) and are neutralized. Atoms can also be injected directly into the cell in a neutral form. In both cases, the vapor of the radioactive atoms sticks to the inner wall of the cell for a period of a characteristic sticking (dwell) time and return to the vapor. Obviously, to ensure the efficient trapping of these short-lived radioactive isotopes, the sticking time of the atoms to the cell wall must be shorter than their radioactive lifetime. To achieve this is important as it will prevent a loss of the atoms that are subject to radioactive decay on the cell wall. Low trapping efficiency can be also being attributed to a large loss through high chemical sorption, when the atoms react chemically with the wall and are irreversibly removed from the vapor.

When a trapping experiment using stable rare atoms injected from a weak source is carried out, a longer sticking time (which can be longer than the filling or escape time of the atoms through the entrance port) leads to a rather long saturation time of both the inner wall and the volume of the cell that leads to long period of time for the creation of sample of trappable atoms in the resonance cell, a step that is so important to commencing an efficient cooling and trapping process.

The interaction of the vapor with a cell wall is also a serious problem in experiments on the light-induced drift (LID) effect. This problem is somewhat different from the trapping of radioactive and rare atoms. In this kind of experiment, resonance atoms can either be pushed or pulled inside or outside a capillary cell by light. When the atoms inside the capillary are pushed by light, they stick to the wall as soon they come. In this experiment, a free passage of the atoms in the capillary, before they stick to the wall, is normally proportionate to the diameter of the capillary. When the sticking time lasts a long time, the fraction of the incoming atoms absorbed on to the wall is much higher than in the vapor phase. As a result, it takes far too long time to saturate the wall with a relatively weak flux of incoming atoms that are pushed by light. The opposite is also true: when the atoms are pulled out of the saturated capillary by light, it takes rather a long time to clean the adsorbed atoms off the wall by pulling them out of the capillary using LID. Thus, the long sticking time leads too long time to achieve a steady-state distribution of the atoms in the cell. This masks the manifestation of the LID effect to a considerable extent. The rapid decay of the number of atoms inside the capillary as the result of high chemical sorption when the incoming atoms are irreversibly removed from the vapor, allows only a very short time span in which atoms can penetrate inside the capillary. In an extreme case, when the chemical adsorption probability of the atom is close to one (i.e. just one collision is sufficient for the atom to be lost) the penetration distance can be just of the same order as the diameter of the capillary.

It is known that as atoms collide with the surface, they undergo an attractive potential whose range depends on the electronic and atomic structures of both the surface and the atoms. Therefore a fraction of the atoms is physadsorbed in the attractive potential well at the surface. Physical adsorption is characterized by an adsorption energy $E$ that determines the sticking time

$$\tau_s = \tau_0 e^{E/kT} \quad (1)$$

where $\tau_0 \sim 10^{-12}$ s. [20].

At the present time, many publications are dealing with studies of different sorts of non-stick coatings with a view to overcoming the problem discussed above. Wieman et al. performed a study of Cs with a dry-film coating on Pyrex to make an efficient atom collection in vapor cell magneto-optical traps in which adsorption energies (0.40±0.03 eV) and sticking time (< 35 μs ) were measured [21]. An attempt to measure the sticking probability or number of bounces of Rb atoms in an MOT Pyrex cell coated with a PDMS compound was performed in [22]. In the paper, it was reported that curing of the coating with Na vapor at a pressure of $10^{-7}$ mbar for 8 days was necessary to obtain a number of bounces of the order of $10^4$. The results of the measurements of the adsorption energy and sticking time for Rb achieved by other authors are summarized in Table I.

TABLE I. The results of the measurements of the adsorption energy and sticking time for Rb.

| Surface materials Ref. | $E$(eV) | Sticking time (s) |
|---|---|---|
| Paraffin coated pyrex Rb [23] | 0.1 | $4 \cdot 10^{-10}$ |
| Tetracontane coated glass Rb [24] | 0.06 | $10^{-11}$ |
| Tetracontane coated pyrex Rb [25] | 0.062 | $10^{-11}$ |

One of the methods employed to measure these quantities is based on a measurement of the time it takes for the test cell to fill with the atoms concerned by opening and closing a valve between the cell and the atomic vapor source. Another method is based on a measurement of the change in the vapor density as the wall temperature is lowered by cooling the cell. Unfortunately, these methods do not allow the measurement of the sticking time and sticking probability with a high enough degree of accuracy and they only offer an approximate value of the quantities that might be obtained. In the problem of the first method was that it was not possible to open or close the valve in a time of less than 0.1 second, therefore it is impossible to ignore the opening and closing times of the valves in the measurements. The second one is only valid when the total number of alkali atoms in the cell is constant. This requirement is difficult to satisfy because of the high chemical activity of the atoms concerned that causes them to decay in the closed cell, irrespective of a change in its temperature.

The work described in this article is focused on measuring the sticking probability and sticking time of Rb atoms on a film made of a polydimethylsiloxane (PDMS) compound. We measured these quantities using Knudsen's flow of Rb atoms in capillary cells. We shall demonstrate that the proposed method is free of the drawbacks encountered in other experiments and allows a more accurate measurement of both the sticking probability and the sticking time to be made. We used cells with a capillary coated with two different types of PDMS coatings of different viscosities. The experimental studies are preceded by a discussion of a model of the atoms in Knudsen's flow in the capillary and by a definition of the relevant parameters..

## II. EXPERIMENT AND THEORETICAL COSIDERATIONS

### A. Experimental set-up

A sketch of the set-up is shown in Fig. 1. The experimental set-up consists of a probe laser -1, a turbo pump -2, a cylindrical glass cell -3, a source of Rb atoms that contains a drop of natural rubidium -4, a glass capillary inside the cell -5, a photodiode -6, a data acquisition n system (DAQ) -7, gauges -8 and 9, photographic flash lamp -10 and a source of Na atoms -11.

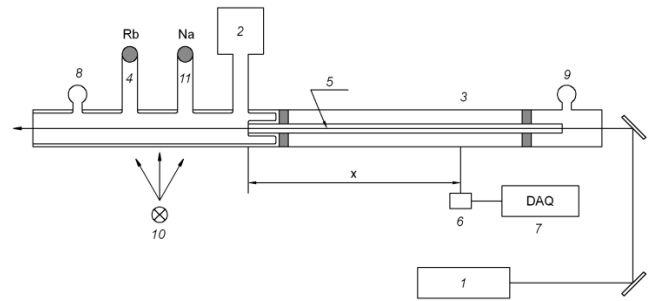

FIG. 1. Experimental setup.

The density of the Rb vapor was measured through the detection of the intensity of the atomic fluorescence by a fast, moveable photodiode connected to a data acquisition system (DAQ). The fluorescence was excited by a free-running diode laser with a frequency tuned to a Rb atom resonant transition of 780 nm. The fluorescent signals were processed by a digital oscilloscope connected to the computer. The DAQ system allows us to collect data with 0.1 ms resolution in time as well as to measure the variation in the Rb vapor over a wide range. The temperature of the cell walls was measured by a digital thermometer. The absolute Rb vapor density at the origin of the capillary and the Rb source is estimated from the temperature of the of Rb metal drop [26].

In the experiment we used two groups of three glass capillaries with diameters of 16 mm, 5 mm and 2 mm in each group. The inner surface of the cell area close to the Rb source and one group of three capillaries were covered by a non-stick coating prepared from a 3% solution of commercial polydimethylsiloxane (PDMS, Mw. 92.400, 11000 mm$^{-2}$s$^{-1}$ viscosity) in ether. Another three capillaries were coated with a different compound (PDMS Mw.182.600, 410000 mm$^{-2}$s$^{-1}$ viscosity). We call the first type, the PDMS with a median viscosity, mv PDMS; the second type with a higher viscosity, hv PDMS. Both types of PDMS were bought from the Aldrich Chemical Company Inc. The cell preparation is described, for example, in [27]. The chosen capillary is held inside the cell by two aluminum perforated disks, which allow the whole cell to be easily pumped but, because of the absence of the coating between these disks, prevent any penetration of the desorbed atoms around the capillary.

The turbo pump provides a vacuum of up to $10^{-7}$ mbar in the cell. The rest gas pressure is measured by two vacuum gauges: one attached to the cell area close to the Rb source and the vacuum pump; the other attached to the entrance of the cell where the probe laser beam enters the cell. We consider a steady-state vacuum is obtained when both gauge readings are stable and the gauge which is kept right away from the pump shows a bit less vacuum than that near the Rb source and the vacuum pump. The steady-state level of the vacuum in a cell with any capillaries inside is usually reached after one week of continuous pumping. This was verified again by an RF discharge switched on inside the capillary. In this the RF voltage is applied to two gauges that are disconnected from their monitors. The indicator that the steady-state vacuum is achieved is that the discharge luminescence is weak, stable and uniform along the length of the capillary.

## B. Theoretical considerations

Let us consider the diffusion of atoms in the evacuated capillary in detail and assume that the atoms collide with the wall only. These atoms move from the initial position along a distance (a one-dimensional shift along the capillary axis)

$$z(t) = \sum z_i \quad (2)$$

The mean quadratic shift is:

$$\langle z^2 \rangle = (\sum z_i)^2 = N \langle z_i^2 \rangle = \frac{t}{\tau} a^2 \quad (3)$$

where $t$ is time, $a^2$ and $\tau$ are the mean quadratic elementary shift and the mean elementary time between two collisions, respectively. On the other hand we know that

$$\langle z^2 \rangle = 2Dt \quad (4)$$

where $D$ is the diffusion coefficient and we can write down:

$$D = \frac{a^2}{2\tau} \quad (5)$$

The time between two collisions $\tau$ consists of the time needed by the atoms to fly between the walls - the mean pass time $\tau_f$ and the sticking time $\tau_s$.

$$\tau = \tau_f + \tau_s \quad (6)$$

therefore, the diffusion coefficient $D$ can be written as:

$$D = \frac{\tau_f D_0}{\tau_f + \tau_s} \quad (7)$$

where $D_0$ is the diffusion coefficient for the case $\tau_s = 0$. In a Knudsen's flow in a cylindrical tube with a diameter $d$ [28]

$$D_0 = \frac{d \bar{v}}{3} \quad (8)$$

and

$$\tau_f = \frac{d}{\bar{v}}, \quad (9)$$

where $\bar{v} = \sqrt{\frac{8kT}{\pi m}}$ is the average atomic thermal velocity at temperature $T$, $k$ – is Boltzmann constant and $m$ is the mass of the atom. For the natural Rb average mass of $85.49 \cdot 10^{-27}$ kg and room temperature of 298 K, $\bar{v} \sim 2{,}7 \cdot 10^4$ cm/s. For three capillaries with diameters of 16 mm, 5 mm and 2 mm used, the mean pass time $\tau_f$ is 59 μs, 18,5 μs and 7,4 μs, respectively.

The diffusion flow of Rb atoms in a capillary without a buffer gas is governed by the following one-dimension diffusion equation:

$$\frac{\delta n}{\delta t} = D\frac{\delta^2 n}{\delta x^2} - \frac{n}{\tau_N}, \tag{10}$$

where $n$ is the density of the atomic vapor and $D$ the diffusion coefficient of the atomic flow in the capillary,

$$\tau_N = N(\tau_f + \tau_s) \tag{11}$$

is the lifetime of the atoms before they are absorbed into the inner surface of the capillary by chemical adsorption, $N$ – the number of bounces (collisions) of the Rb atoms in the capillary before their chemical reaction to the PDMS coating commences. The sticking probability of an Rb atom to the coating is a value $1/N$. It varies from zero to one.

The steady-state solution ($\delta n/\delta t = 0$, no flash light can be written in the following simple form:

$$n(x) = n_0 \exp(-x/l), \tag{12}$$

where $n_0$ is the vapor density at the origin of the capillary ($x = 0$) and

$$l = (D\tau_N)^{1/2} \tag{13}$$

is the characteristic length of the decay in density along the capillary. Now by using equations (3), (4) and (11) we can write:

$$N = \frac{3l^2}{2d^2} \tag{14}$$

Then, taking into account of the characteristic length $l$ measured in the experiment and the diameter $d$ we can calculate the number of bounces (collisions) made by an Rb atom in the capillary before its adsorption by the PDMS coating and the sticking probability of the Rb atom to the PDMS coating.

In order to obtain the sticking time $\tau_s$, we have to solve non-steady-state equation (10) with the initial condition

$$n_{t=0} = \delta(0) \tag{15}$$

and the boundary condition

$$n \to 0 \text{ when } x \to \infty \tag{16}$$

where $\delta x$ is a delta function which describes a burst of photo-desorbed atoms by the photographic flash at the capillary origin with $x = 0$. By using a trivial solution, we find a solution $n = n(x, t)$ that satisfies both the initial and the boundary conditions:

$$n(x,t) = \frac{A}{2\sqrt{\pi D t}} \exp\left[-\frac{x^2}{4Dt}\right] \tag{17}$$

where $A$ some constant. There are two possibilities to find the sticking time $\tau_s$. Using equation (17) it is possible to find $t_m$ when the density of the atom n approaches its maximum

$$t_m = \frac{x^2}{2D} \tag{18}$$

Then we can write down

$$\tau_s = \tau_f\left(\frac{D_0}{D} - 1\right), \tag{19}$$

where $D$ can be found from the equation (18), using the experimental values of tm and $D_0$ given in equation (8). Another way to obtain the sticking time $\tau_s$ is to measure a delay time $t_d$ – a time at the flex point, i.e. where the second derivative of the fluorescence curve is equal to zero. Using equation (17), we can evaluate $D$ as:

$$D = 0{,}0459 \frac{x^2}{t_d} \tag{20}$$

and using the experimental value of $t_d$, x and equations (8), (19), we obtain the sticking time $\tau_s$. In the following measurements and calculations of the sticking time, we use the second possibility only. Note, that in these calculations we ignore the term $n/\tau_N$ in equation (10). This omission is possible because the lifetime of the Rb atoms in the cell $\tau_N$ is much longer than either the $t_m$ or $t_d$.

## III. RESULTS AND DISCUSSIONS

We found that a freshly coated cell did not show any fluorescence from the Rb atoms, meaning that the lifetime and the number of bounces of the atoms in the cell were both very small. This can be attributed to the fact that the fresh coating in the cell will have had a chemically active surface and volume, probably because of traces of a gas like oxygen or of water adsorbed and mixed with the molecules in the coating. To minimize the residual chemical activity of the coating, we carried out a curing (or passivation) procedure [5, 22, 29] using Rb. Maintaining a continuous pumping of the cell, we heated up the source of the Rb atoms so that the pressure of the alkaline vapor in the cell was about $10^{-7}$ mbar. After a few days of this curing, fluorescence appeared; first near the source and origin, then slowly spreading as far as the entrance of the capillary. After one week of the curing process at the higher temperature, the pressure of the Rb vapor was reduced by keeping the source at room temperature. After the curing the fluorescence became more or less uniform along the length of the cell. The steady-state distribution of the Rb vapor along the capillary could be typically achieved after one or two weeks of exposure to the Rb vapor, but it still never reached a uniform distribution along the capillary. We believe that the observed decay in the density of the Rb vapor in the completely passivized capillary is determined by both the irreversible chemical reaction of the atoms with the PDMS coating and by the slow diffusion of the Rb atoms into the coating towards the glass substrate [23]. We found that a cleaning of molecular impurities from the area close to the Rb source and the origin of the capillary by the illumination of this area by a powerful halogen lamp can improve the vacuum and the reproducibility fluorescence dynamic [30].

Because we measure the density of the Rb atoms by the detection of the atomic fluorescence intensity, an optical pumping through the hyperfine atomic levels of the atoms can deform the fluorescence signal and hence distort the dynamics of the photo-desorbed atoms. To avoid the influence of the optical pumping before any experiment begins, we reflect a small amount of the laser radiation back to the laser. In reaction to this feedback, the laser starts to generate a spectrum consisting of a central large peak with two small sidebands separated by about 3 GHz from the central peak. This spectrum is generated by the so-called "oscillation relaxation effect" in diode lasers (additional information can be found in [31]). Then we tune the central peak to the maximum of $5S_{1/2}$, F = 3 – $5P_{3/2}$, F' = 4 $^{85}$Rb transition, while one of the side bands automatically tunes to $5S_{1/2}$, F = 2 – $5P_{3/2}$, F' = 3 $^{85}$Rb re-pumping transition. This simple method allows us to illuminate the optical pumping effect in our experiment completely without having to use a re-pumping laser. It was verified that the signal made by suddenly opening the laser beam demonstrates a perfectly rectangular form.

The measurement of the number of collision $N$ is as the following: keeping the cell pumped at a constant temperature we heat the Rb vapor source up to a temperature of $300^0$C. This creates a density in the Rb vapor of $1.6 \cdot 10^{10}$ cm$^{-3}$ that is high enough to allow us to detect clearly a steady-state distribution of the vapor along the capillary, but low enough not to cause the deterioration of the vacuum in the cell. Then we detect the distribution of the Rb density along the capillary and measure the characteristic length $l$. A typical result of the steady-state fluorescence intensity as a function of the distance x is shown in Fig. 2. The graph is taken from a 2 - mm diameter capillary coated with hv PDMS after two months' of pumping.

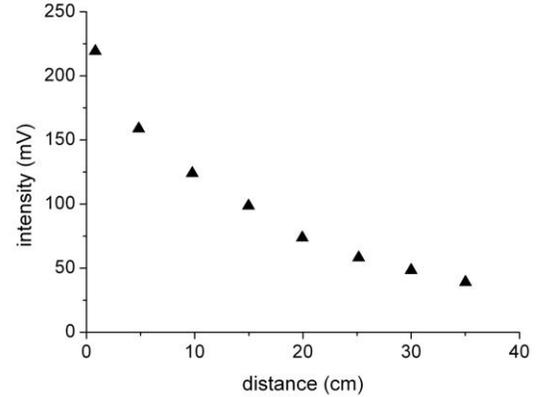

FIG. 2. Steady-state fluorescence intensity of Rb atoms as a function of the distance from the capillary origin. The graph has been derived from a 2 – mm diameter capillary coated with hv PDMS.

In Fig. 2 it can be seen that the density of the Rb vapor reaches 1/e-level at a distance $l$ of 25 cm from the capillary origin. From equation (13) and with $l$ = 25 cm, $d$ = 2 mm it is possible to calculate $N$. It was found that it takes $2.3 \cdot 10^4$ collisions (sticking probability $4.3 \cdot 10^{-5}$) for the Rb atoms to be irreversibly absorbed into the hv PDMS coating. This result is consistent with the values published in the literature for various alkalis, preponderantly Na on paraffin passivized by Na atoms [13] and Rb on PDMS passivized by either Na or Rb atoms [22]. Results of the measurements of N as a function of the pumping time are reported in Table II for 2-mm-diameter capillaries coated with hv PDMS.

TABLE II. Number of collisions of N as a function of pumping time for 2-mm-diameter capillaries coated with hv PDMS.

| Pumping time | $N$, number of collisions |
|---|---|
| 1 week | $\sim 10^4$ |
| 1 month | $2 \cdot 10^4$ |
| 2 months | $2,3 \cdot 10^4$ |

In principle, the fairly long vapor distribution and large number of bounces observed can be attributed to the collision of the Rb atoms with the film surface at the detection point that release by sputtering of other Rb atoms previously stored at this point. This exchange process was found by Guckert et al. [32] be important for radioactive atoms, whereas it is not noticed a stable atoms experiment. We checked the negligibility this possible systematic effect by curing the capillary using Na atoms that cannot be detected by the probe laser light used. We found that the fluorescence behavior and the number of bounces were exactly the same as for Rb passivized cell.

We perform the measurement of the sticking probability as a function of the PDMS film temperature. To do this, we cool down the part of the cell containing the capillary using liquid nitrogen and then leave the cell to heat itself up until room temperature is reached, then measure l and calculate the sticking probability. To take measurements above room temperature, we heat up the same cell area with the capillary using a heater. Measurements of the sticking probability of the Rb atoms to hv PDMS as a function of temperature are presented in Fig. 3, taken after 2 months' pumping. It is evident that in a temperature range from -$100^0$C to $160^0$C the sticking probability does not change to any great extent, but it does sharply increase at a temperature lower than -$100^0$C. This temperature corresponds to the surface adsorption energy of E = 0.026 eV, that implies, according to the equation (1), a sticking time $\tau_s$ of $1.4 \cdot 10^{-12}$ s.

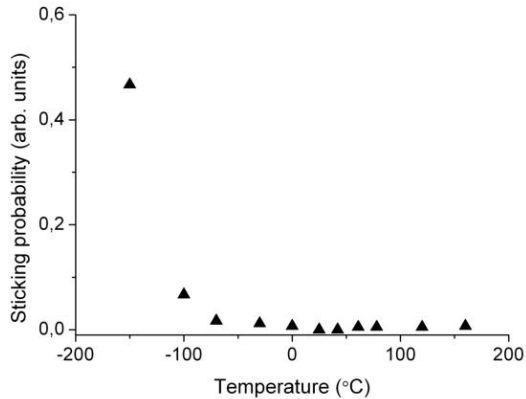

FIG. 3. Sticking probability $1/N$ for the adhesion of Rb atoms to hv PDMS as a function of temperature of the PDMS film.

To measure directly the sticking time of Rb atoms to the PDMS film, we cool down of the Rb vapor source to lower the steady-state density of the vapor. Then we illuminate the area of the cell (as shown on Fig. 1) close to the Rb source and the origin of the capillary by a photographic flash lamp. Next we detect the dynamic of the fluorescence signal of the burst of desorbed atoms and measure the delay in the arrival of the atoms at the detection point. We measure the delay time as a function of the distance between the detector and the origin. In this ultra-high vacuum condition, the delay in the arrival of the atoms at the detection point is principally attributable to the many collisions of the atoms with the capillary wall and this delay can be increased by a non-zero stick time in every collision. Because the mean path time in a narrow capillary is relatively small, even a very short sticking time will have a significant influence on the delay.

Fig. 4 shows two fluorescence curves taken at a distance of 23 cm from the origin of two capillaries: one 16 mm (curve A) and the other 2 mm (curve B) in diameter. In both cases, the coating is hv PDMS. It can be seen that the ratio of $t_d^B/t_d^A$ and $t_m^B/t_m^A$ is also ~ 8, in accordance to equations (18) and (20), that predict a dependence on the capillary diameter d. This result can be considered a check on the agreement between the theory and the experiment.

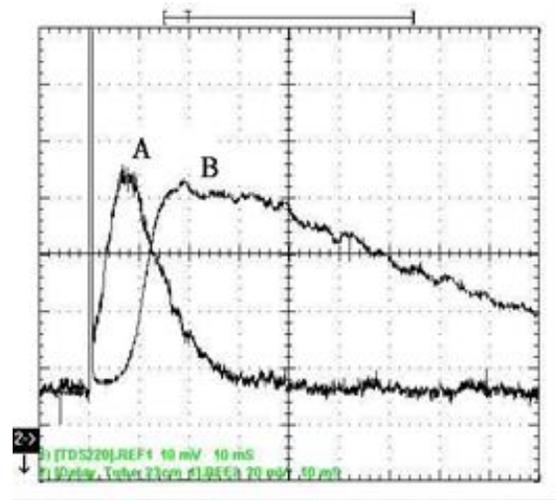

FIG. 4. Fluorescence intensity of Rb atoms as a function of time triggered by flash light. In curve A, the capillary diameter is 16 mm, while in curve B the capillary diameter is 2 mm. In both cases the coating is the hv PDMS and the signals have been taken at a 23 cm distance from the capillary origin.

Fig. 5 shows a typical example of a recording of the Rb fluorescence intensity as a function of the time taken in a cell with a capillary of 2 mm coated with mv PDMS compound. This curve was taken at a distance of 20 cm from the origin. It can be seen that the intensity of the fluorescence of the Rb atoms as a function of the time immediately after the flash (sharp peak at $t = 0$) is close to the low steady-state intensity at a selected distance of the detector from the capillary origin. After some delay, it increases and then, after approaching of a maximum, the intensity slowly declines the beginning intensity again. Note that the form of this curve is in complete accordance with the prediction of equation (17).

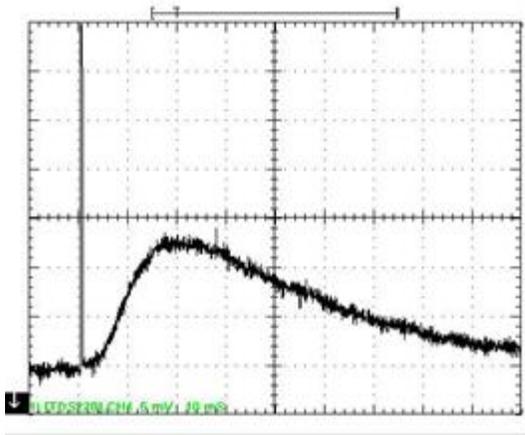

FIG.5. Fluorescence intensity of Rb atoms as a function of the time triggered by a flash of light in the 2 mm capillary cell coated with mv PDMS. The time scale is 10 ms/div, while the intensity scale is 5 mV/div.

Measurements of $t_d$ as a function of the distance $x$ are reported in Fig. 6 for 5-mm-diameter capillaries coated with two sorts of PDMS. For hv PDMS the data were collected in three periods: one week after installation, after one month and after two months of pumping. The data for mv PDMS were taken after two months' pumping. The theoretical curve for $\tau_s = 0$ s that calculated above for the Rb atoms is also shown. From these measurements it can be seen that td is proportional to $x^2$ as predicted by the equations (20).

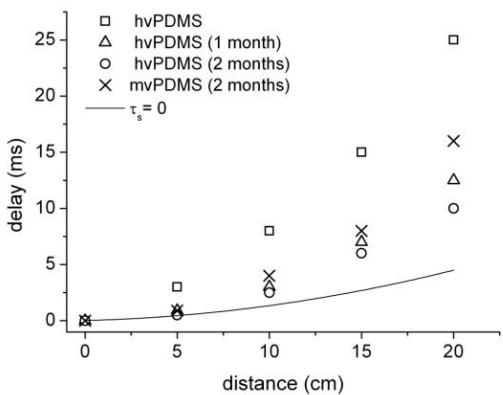

FIG.6. Measurements of the delay td as a function of the distance x between the capillary origin and the detector. The capillary diameter is 5 mm. Data for the mv PDMS and the hv PDMS at different times and the theoretical curve for $\tau_s = 0$ s are reported. Error bars have been omitted for the sake of clarity.

In order to calculate the sticking time $\tau_s$, we derived our data from equation (20). We evaluated the uncertainty $\delta\tau_s$ using the standard deviation given by the fit for the values of $D$. The calculated sticking times are reported in Table III.

TABLE III. Sticking time and its uncertainty calculated by fitting the data in Fig. 6 according to the theoretical model

| Coating | $\tau_s$(μs) | $\delta\tau_s$(μs) |
|---|---|---|
| hv PDMS | 111 | 10 |
| hv PDMS (1 month) | 38 | 4 |
| hv PDMS (2 months) | 22 | 3 |
| mv PDMS (2 months) | 49 | 6 |

It can be seen that the experimental results tend to approach the theoretical ones only after a long-term pumping of the coated cell. For example, the results shown in Fig. 6 demonstrate that the deviation between the experimental and the theoretical td becomes smaller if the sort of coating is changed from mv PDMS to hv PDMS and/or if the cell is pumped for a longer time. For example, after two months' pumping, the difference between the experimental and theoretical curves for hv PDMS grows smaller, and this is important, but not a zero.

Note, that the measured times for both types of PDMS is about million times larger than the sticking time derived above from the surface adsorption energy of Rb atom and temperature of the coating that is of $2 \cdot 10^{-11}$ s. These times are consistent with measured in [21] for Cs atom on a dry-film coating on Pyrex (<35 ms) and they are 10 times larger than in [33] for Rb atom for OTS- and paraffin-coated surfaces that are 0.9±0.1 μs and 1.8±0.2 μs, respectively.

It is clear that this surprisingly large difference between the sticking time estimated from the equation (1) and the sticking time measured directly in the non-steady-state experiment cannot be attributed to a deceleration of the Rb atoms as a result of their collision with the rest gas molecules. In fact, the magnitude of the mean free path of the Rb atom in the rest gas $l_{path}$ is as the following:

$$l_{path} = (Sn)^{-1}, \qquad (21)$$

where $S$ is the effective cross section for the collision of Rb atoms with the rest gas molecules and n is the density of these molecules. At a working vacuum pressure of $10^{-7}$-$10^{-6}$ mbar, $n$ is in the order of $10^9$- $10^{10}$ cm$^{-3}$. Taking into account this estimated $n$ and the typical cross section for a collision in the order of $10^{-15}$ cm$^2$, we find that $l_{path}$ is in the order of $10^6$- $10^5$ cm. Hence, the estimated mean free path of the Rb atoms in the cell is much larger than the diameters and lengths of all the capillaries used in our experiments. Therefore the influence of the collision of Rb atoms with the rest gas molecules on the results of the measurements is surely negligible.

A possible interpretation of the difference can be argued as follows. When atom collides with a surface it can be linked there some period of time because of Van der Waal's attractive force. We believe that once trapped, the atom can diffuse inside a PDMS film, then diffuse back to the surface and then, in turn, be desorbed back to the vapor. Hence, a long sticking time has to depend on how deeply the atoms can diffuse inside the coating. This is suggested by the fact that the sticking time for mv PDMS that has a higher diffusion rate has a longer sticking time than hv PDMS, in which the diffusion is less pronounced. It is known that PDMS films exhibit slow decreasing permeability and diffusion rate with long aging time [34]. This can explain slow decreasing of the sticking time during two months' that has observed in the experiment. Note that the adsorbed atoms can be attracted inside the film by the coating convective flow.

We believe that choosing a surface that has a low diffusion rate (or, in other words, has a high viscosity) will make it less likely that the atoms will remain in the coating for too much long period of time. Probably, that the equation (1) is valid for particles collision with not liquid but truly solid surface only.

## IV. CONCLUSIONS

We present the results of a systematic study of Knudsen's flow of Rb atoms in cylindrical cells (tubes and capillaries) coated with two types of polydimethylsiloxane (PDMS) compounds of different viscosities. We have developed a one-dimensional model that agrees with our observations of the diffusion flow of Rb atoms in a capillary. Our model includes the decay of the density of the Rb vapor along a capillary because of the chemical adsorption of atoms into the PDMS film surface, the dependence of the delay of the desorbed atoms at the detection point on both the mean free pass time and the sticking time. We have shown that, because of the many collision of the Rb atoms with the capillary under ultra-high vacuum conditions, it is possible to estimate accurately both the sticking probability and the sticking time. We have described the results of the measurements of these parameters for two types of PDMS of different viscosities and as a function of the physical aging of the organic film. We observed a fairly long sticking time and propose a tentative explanation for this surprising result.

## ACKNOWLEGEMENTS

We would like to thank Z. Peshev for the help in the experiment, N. Surovtsev for fruitful discussions and V. Sorokin for the careful reading of the manuscript.

*Corresponding author: atutovsn@mail.ru